# Nearly constant loss – the 2nd universality of AC conductivity by scaling down subsequent random walk steps by, 1/sqrt(t)

Baruch Vainas

The Weizmann Institute of Science, Rehovot 76100,  Israel

Abstract

In the frequency domain, the nearly constant loss (NCL), is characterized by a slope 1 in log of the real part of the electrical conductivity vs log frequency plots. It can be explained by an anomalous diffusion, defined by a random walk with the mean square displacement proportional to the logarithm of time, rather than being linearly proportional to time, as in normal diffusion. The present work suggests a random walk algorithm that leads to anomalous, logarithmic time dependence. That has been accomplished by scaling down the subsequent random walk displacements by a factor, 1/sqrt(t)

## Introduction

The nearly constant loss (NCL) is a widely observed phenomenon in disordered materials, where the AC electrical conductivity increases nearly linearly with the frequency of the applied voltage [1]. This characteristic is also known as the 2nd universality, given its ubiquitous presence in many different systems, and, following a related, widely observed frequency response in the form of a power law [2, 3], with a positive exponent <1, which is also known as the first universality.

NCL appears over a wide range of frequencies at very low temperatures for which the DC is not significant.  At high temperatures it could be observed at high frequencies only.  While the first universality is a thermally activated ion hopping process, the NCL phenomenon is not thermally activated [4]. The low temperature NCL and the one observed at high temperatures and high frequencies are suggested to have different origins, and it is an open topic [1].

It has been shown that a logarithmic dependence of the mean square displacement (MSD) of a particle on time as it performs a random walk from the point of origin can be translated to a NCL in the frequency domain [5, eq. 10] for mobile ions in disordered solids.

While it is quite easy to model the classic diffusion process leading to, MSD ~ t, where "t" is time, by a random walk simulation algorithm, an algorithm leading to, MSD ~ log (t), is not trivial to derive and, to the best of author's knowledge, it does not appear in literature.

In the following section a "slowed down" random walk algorithm is shown to lead to a logarithmic dependence on time, which could contribute to a better understanding of NCL, which is still not fully understood and is actively studied.

## Results and Discussion

The initial implementation of the algorithm suggested here, simulates a simple random walk leading to normal diffusion, MSD ~ t characteristic, with unit time step, and a constant absolute value of the displacement for each time step. The direction of the hopping is random, with 6 degrees of freedom. For say, the X-axis, we have +delta(x), or -delta(x), only. It was found that an addition of small probability of a "no move" degree of freedom (for a "lazy" walk) did not change the results significantly.

The algorithm calculates the 3D position of the particle at a discrete time "t", expressed as the number of time steps from the start of the random walk run, for many realizations of the random walk. All 3-axis components of displacements between the initial and the position at time "t" for each realization of the random walk, are squared, added, and then divided by the number of runs giving the mean square displacement value (the MSD) for all realizations after the same number of time steps.

The first experiment was to simulate a classic random walk that is expected to result in, MSD ~ t.

In this simulation:

a. total number of time steps in a random walk run = 2000

b. single displacement absolute value for a time step = 1

c. probability of +/0/- hops is, 8/20, 4/20, 8/20 respectively (see line 6 from top in the Appendix)

d. total number of random walk runs =2000

The X-axis component line in the algorithm, that is making the recursive addition of the next time step displacement is:

$X_n \leftarrow X_{n-1} + step*S1_{n-1}$ (the case of normal random walk, not shown in the appendix),

where step=1, $S1_{n-1}$ can have +1/0/-1 values determined by probabilities mentioned in "c" in the list above. This leads to the normal random walk.

As can be seen in Fig. 1, below, MSD vs time steps, in linear scales, shows the expected straight line, characteristic of classic diffusion, with the slope proportional to the diffusion constant.

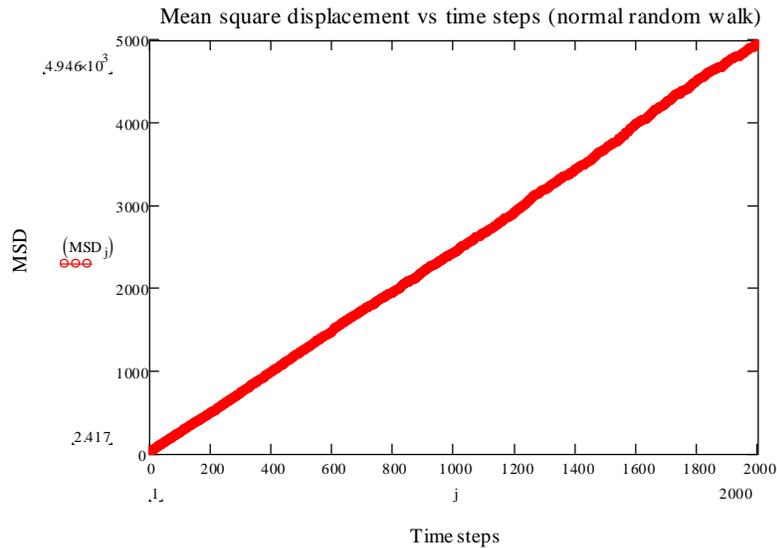

Fig. 1: Mean squared displacement as a function of time (in units of time steps) for normal random walk

For the second experiment, the size of a single time step displacement was modified to be a function of the discrete time – the number of the time step in the sequence of the random walk run. All other parameters are the same as in the first experiment. The modified recursive addition line in this case (shown in the appendix, code line 4 from bottom), is:

$X_n \leftarrow X_{n-1} + (step/sqrt(n))*S1_{n-1}$ (the slowed-down random walk),

where "n" is the discrete time – the time step number in the sequence of steps.

This results in the apparently logarithmic dependence of the mean square displacement on time, as shown in Fig. 2:

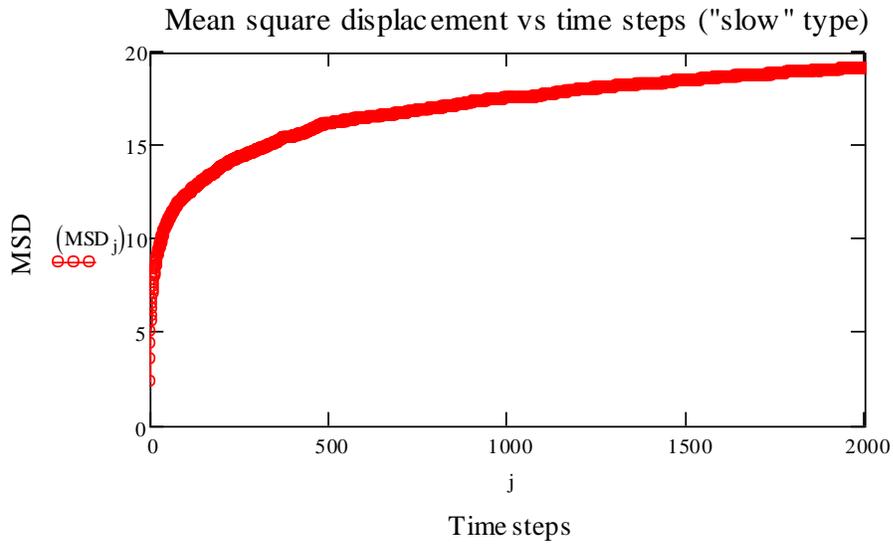

Fig. 2: Displacement at discrete time, t is reduced by the factor 1/sqrt(t).

Transforming the data in Fig. 2 to semi logarithmic format (logarithmic time scaling), gives the expected straight line of MSD ~ log(t).

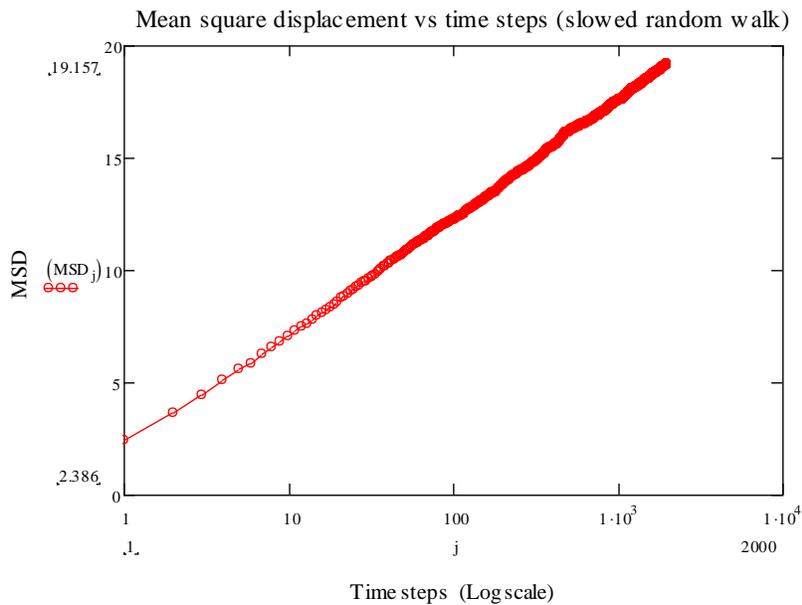

Fig. 3: The slowed down random walk of Fig. 2 in semi logarithmic representation, with log time scaling.

It should be noted that the recursive addition, $X_{n-1} + (step/sqrt(n))*S1_{n-1}$ , used in the present algorithm is a special (p=1/2) case of random hyperharmonic series, for example:

$1 + 1/2^p - 1/3^p + 1/4^p - 1/5^p - 1/6^p \ldots +/- 1/n^p$                eq. 1

For p = 1 and non-random, but alternating +/- signs, the series converges [6]:

$1 - 1/2 + 1/3 - 1/4 + 1/5 - \ldots = ln(2)$                eq. 2

Given the obvious non-convergence shown in Figs. 2-3, the random, p =1/2, hyperharmonic series in our case can not be related to the alternating, p=1, hyperharmonic series. Apparently, after realization of many different random sequences, and using squaring and averaging operations by the algorithm used, we get logarithmic characteristics similar to a simple harmonic series. While a mathematical proof, or explanation, for the clear logarithmic characteristic shown by the algorithm suggested here, is obviously an interesting mathematics topic, it is not in the scope of this note, but one can point to the similarity with random hyperharmonic series, modified by additional operations, of taking squares and averaging. The physical significance of the scaling-down algorithm given here might provide a clue to possible energy dissipation mechanisms that will be researched in more studies.

## References

[1] Dieterich, W. ; Maass, P. Non-Debye relaxations in disordered ionic solids. *Chemical Physics* **2002** *284*, 439–467.

[2] Vainas, B.; Almond, DP.; Luo, J.; Stevens, R. An evaluation of random RC networks for modelling the bulk ac electrical response of ionic conductors. *Solid State Ionics* **1999** *126*, 65–80

[3] Almond, DP.; Vainas, B. The dielectric properties of random *R–C* networks as an explanation of the 'universal' power law dielectric response of solids. *Journal of Physics: Condensed Matter* **1999** *11*, 9081–9093

[4] Funke, K.; Banhatti, RD.; Badr, LG.; Laughman, DM.; Jain, H. Toward understanding the second universality—A journey inspired by Arthur Stanley Nowick. *J Electroceram* **2015** *34*, 4–14

[5] Sidebottom, DL. Understanding ion motion in disordered solids from impedance spectroscopy scaling. *Reviews Of Modern Physics* **2009** *81*, 999-1014

[6] Hudelson, M. Proof Without Words: The Alternating Harmonic Series Sums to ln 2. *Mathematics Magazine* **2010** *83,* 294–294

## Appendix

The main part of the algorithm suggested, written in the programming module of Mathcad, version 13.0 (Mathsoft)

$DD_{NN, times} \leftarrow 0$

$W \leftarrow 10$

for $i \in 1 .. times$

    $r1 \leftarrow runif(NN, -W, W)$

    $r2 \leftarrow runif(NN, -W, W)$

    $r3 \leftarrow runif(NN, -W, W)$

    for $k \in 0 .. NN - 1$

        $S3_k \leftarrow if(r3_k < Q, -1, if(r3_k > P, 1, 0))$

        $S2_k \leftarrow if(r2_k < Q, -1, if(r2_k > P, 1, 0))$

        $S1_k \leftarrow if(r1_k < Q, -1, if(r1_k > P, 1, 0))$

    $Z_0 \leftarrow 0$

    $Y_0 \leftarrow 0$

    $X_0 \leftarrow 0$

    $PP_0 \leftarrow 0$

    for $n \in 1 .. (NN)$

        $Z_n \leftarrow Z_{n-1} + \dfrac{step}{\sqrt{n}} \cdot S3_{n-1}$

        $Y_n \leftarrow Y_{n-1} + \dfrac{step}{\sqrt{n}} \cdot S2_{n-1}$

        $X_n \leftarrow X_{n-1} + \dfrac{step}{\sqrt{n}} \cdot S1_{n-1}$

        $PP_n \leftarrow (Z_n)^2 + (X_n)^2 + (Y_n)^2$

    $DD^{(i)} \leftarrow \overrightarrow{(DD^{(i-1)} + PP)}$

$DD^{\langle times \rangle} \leftarrow \dfrac{DD^{\langle times \rangle}}{times}$

Zeroing all displacement data - for all times steps and runs. The total of time steps (NN) per realization of a random walk - a run, is typically 2000. The number of realizations - runs (times) is typically 2000 as well

3 vectors: for all NN time steps in 3D runs, i, random numbers from a uniform distribution between -W and +W (typical W=10) are assigned for probability calculation in the next step

Probabilities for displacements in 3D: for X-axis, probabilieties for left/right/none, etc, for all time steps, k, of total NN. For division of the +/-W domain into 3 regions, typically, P= 2, Q= -2. Probabilities determine corresponding S parameter to be 1, -1, or 0

Initialization of each, i, run for the initial position in 3D

Recursive addition of possible displacements in 3D given by the 3 S values (1, -1, 0) for each time step, n, of NN for a given run, i, multiplied by fixed time step displacement, "step", set by default to 1, and in the case of "slowed down" random walk, also multiplied by the scaing down factor 1/sqrt(n), "n" being the discrete time, t, or the number of the time step

Square displacement in 3D, for a run, i, for all NN time steps

Summation of all square displacements for runs, i, their total number = "times", typically 2000.

Mean square displacement (MSD)